\documentclass[a4paper,preprintnumbers,floatfix,nosuperscriptaddress,pra,twocolumn,showpacs,notitlepage,longbibliography]{revtex4-2} 

\usepackage[utf8]{inputenc}  
\usepackage[T1]{fontenc}     
\usepackage[english]{babel}  

\usepackage[colorlinks,citecolor=blue,linkcolor=magenta,urlcolor=blue]{hyperref}  
\usepackage{amssymb,amsthm,mathtools,dsfont}
\usepackage{physics}
\usepackage[capitalise,noabbrev]{cleveref}			
\usepackage[caption=false]{subfig}

\newtheorem{protocol}{Protocol}

\DeclareMathOperator{\1}{\mathds{1}}

\DeclareMathOperator{\AAA}{\mathcal{A}}
\DeclareMathOperator{\EE}{\mathcal{E}}

\DeclareMathOperator{\cnot}{\textsc{cnot}}

\begin{document}
\title{Entanglement Purification of Hypergraph States} 

\author{Lina Vandr\'e}
\affiliation{Naturwissenschaftlich-Technische Fakult\"at, Universit\"at Siegen, Walter-Flex-Stra\ss e 3, 57068 Siegen, Germany}
\author{Otfried G\"uhne}
\affiliation{Naturwissenschaftlich-Technische Fakult\"at, Universit\"at Siegen, Walter-Flex-Stra\ss e 3, 57068 Siegen, Germany}

\date{\today}

\begin{abstract}
Entanglement purification describes a primitive in quantum 
information processing, where several copies of noisy 
quantum states are distilled into few copies of nearly-pure 
states of high quality via local operations and classical 
communication. Especially in the multiparticle case, the task of entanglement 
purification is complicated, as many inequivalent forms 
of pure state entanglement exist and purification protocols 
need to be tailored for different target states. In this 
paper we present optimized protocols for the purification 
of hypergraph states, which form a family of multi-qubit states
that are relevant from several perspectives. We start by
reformulating an existing purification protocol 
in a graphical language. This allows for systematical 
optimization and we present improvements in three directions. 
First, one can optimize the sequences of the protocol with 
respect to the ordering of the parties. Second, one can
use adaptive schemes, where the measurement results obtained
within the protocol are used to modify the protocols. Finally, 
one can improve the protocol with respect to the efficiency, 
requiring fewer copies  of noisy states to reach a certain target
state. 
\end{abstract}

\maketitle

\section{Introduction}

For many tasks in quantum information processing one needs 
high-fidelity entangled states, but in practice most states 
are noisy. Purification protocols address this problem and 
provide a method to transform a certain number of copies 
of a noisy state into single copy with high-fidelity. 
The first protocols to purify Bell states were introduced 
by Bennett \textit{et al.}\ and Deutsch \textit{et al.}\  \cite{Bennett_1996, Bennett_1996-09, Deutsch_1996}. 
The concept was then further developed for different 
entangled states, especially in the multiparticle setting.  
This includes protocols for the purification of different
kinds of states, such as graph states \cite{Aschauer_2005, Kruszynska_2006}, or W states \cite{Miyake_2005}, see also \cite{Duer_2007} for an overview.
There are several examples for experimental realisations of purification protocols \cite{Kwiat_2001,Pan_2003,Hu_2021,Huang_2022}.

When analysing multiparticle entanglement, the exponentially 
increasing dimension of the Hilbert space renders the discussion
of arbitrary states difficult. It is therefore a natural strategy
to consider specific families of states which enable a simple
description. Graph states \cite{Hein_2004} and
hypergraph states \cite{Kruszynska_2009, Qu_2013, Rossi_2013} form 
such families of multi-qubit quantum states, as they can be described
by a graphical formalism. Besides this, they found applications
in various contexts, ranging from quantum error correction \cite{Shor_1995, Wagner_2018}, measurement-based quantum computation \cite{Raussendorf_2001,Gachechiladze_2019}, Bell nonlocality  
\cite{Scarani_2005,Guehne_2005,Mari_2016},
and state verification and self-testing \cite{Morimae_2017,Baccari_2020}. 
Consequently, their entanglement properties 
were studied in various works \cite{Guehne_2014, Zhou_2020}.
Note that hypergraph states are a special 
case of the so-called locally maximally entangleable  states \cite{Kruszynska_2009}. 

Concerning entanglement purification, the only known purification 
protocol which is valid for hypergraph states is formulated for locally maximally entangleable
(LME) states by Carle,  Kraus, D\"ur, and de Vicente (CKDdV)  \cite{Carle_2013}. In this paper we first ask how this protocol
can be translated to the hypergraph formalism. Based on this, 
we can then systematical develop improvements of the protocol.

Our paper is organized as follows. 
In \cref{sec:hypergrstates} we introduce our notation 
and review hypergraph states. We also  recall how operations 
like $\cnot$ and Pauli operators act graphically. 
In \cref{sec:protocol} we reformulate the CKDdV purification 
protocol  in a graphical manner, providing a different language 
to understand it. Based on this,  we propose systematic extensions in \cref{sec:improvedprot}, which naturally arise from the graphical formalism. We first propose two approaches to make the protocol 
applicable to noisy states where the original CKDdV protocol fails. 
Later we propose a method to requiring fewer copies of noisy states 
to reach a certain target state. In \cref{sec:morequbits} we extend 
the protocol to more qubits. We summarize and conclude in \cref{sec:conclusion}.

\begin{figure}[t]
    \centering
    \includegraphics[width=0.8\columnwidth]{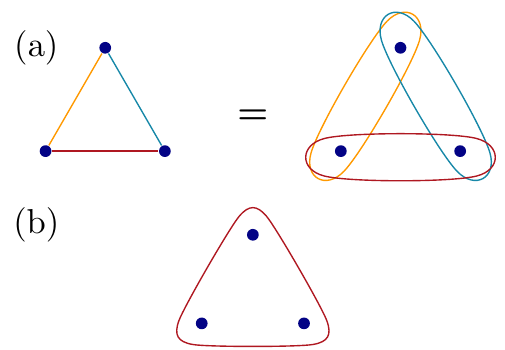}
    \caption{Examples of graphs and hypergraphs. Figure (a) 
    shows a fully connected graph, which corresponds to the 
    three-qubit GHZ state. In the hypergraph state formalism 
    one often draws edges by circles (right) instead of lines 
    as in the graph state formalism (left). The hypergraph 
    state corresponding to the hypergraph in the lower figure (b) 
    of the figure is local unitary equivalent to the state 
    $\ket{H}= \left(\ket{000} + \ket{001} + \ket{010} + \ket{111} \right)/2$. 
    }
    \label{fig:simple_hgraph}
\end{figure}

\section{Hypergraph States} \label{sec:hypergrstates}
In this section we present a short introduction to the 
class of hypergraph states and the description of 
transformations between them. Readers familiar with the 
topic may directly skip to the next section. 

\subsection{Definition of Hypergraph States}
A hypergraph $H = (V,E)$ is a set $V$ of vertices and 
hyperedges $e \in E$ connecting them. Contrary to a 
normal graph, the edges in a hypergraph may connect more
than two vertices; examples of hypergraphs are given in \cref{fig:simple_hgraph}. 
Hypergraph states are multi-qubit quantum states, where the 
vertices and hyperedges of the hypergraph $H = (V,E)$ represent 
qubits and entangling gates, respectively. The state $\ket{H}$, corresponding to a hypergraph $H = (V,E)$ is defined as
\begin{align}
    \ket{H} = \prod_{e \in E} C_e \ket{+}^{\otimes \lvert V \rvert } \equiv U_{\text{ph}} \ket{+}^{\otimes \lvert V \rvert }, 
    \label{eq:hypergrstate}
\end{align}
where $C_e$ is a generalized controlled-$Z$ gate, acting on qubits in the 
edge $e$ as $C_e = \1_e - 2 \dyad{11 \ldots 1}_e$. If an edge 
contains only a single vertex, $\lvert e \rvert = 1$, then 
$C_e$ reduces to the Pauli-Z operator, and for two-vertex edges
$C_e$ is just the standard two-qubit controlled phase gate. 
A detailed discussion on hypergraph state properties can be 
found in Refs.~\cite{Guehne_2014, Mari_2019_Dissertation}. 

Similarly  as for graph states, there is an alternative definition 
using so-called stabilizing operators. First, one can define for 
each vertex $i$ a stabilizer operator, 
\begin{align}
S_i = U_{\text{ph}}  X_i  U_{\text{ph}}^\dagger, 
\end{align}
where $X_i$ denotes the first Pauli matrix  acting on
the $i$-th qubit and $U_{\text{ph}}$ denotes the collection
of phase gates as in Eq.~(\ref{eq:hypergrstate}). Note
that here only the gates with $i \in e$ matter. The stabilizing
operators are non-local hermitian observables with eigenvalues 
$\pm 1$, they commute and generate an abelian group, the 
so-called stabilizer.

Then, a hypergraph state may be defined as a common eigenvector
of all stabilizing operators $S_i$. Here, one has to fix the 
eigenvalues of the $S_i$. Often, the state defined in \cref{eq:hypergrstate} is called  $\ket{H_{00\ldots 0}}$, as 
it is a common eigenstate of the $S_i$ with eigenvalue $+1$.
By applying Pauli-Z gates on the state, one obtains states 
orthogonal to $\ket{H_{00\ldots 0}}$, where some of the eigenvalues
are flipped to $-1.$ By applying all possible combinations of 
Z gates, one obtains a basis: $\lbrace \ket{H_\mathbf{k}} = Z^\mathbf{k} \ket{H_\mathbf{0}} \rbrace$, where $\mathbf{k}$ is a binary 
multi-index and $Z^\mathbf{k} = \bigotimes_{v \in V} Z^{k_v}_v$. 
In this notation, it holds that $S_i \ket{H_\mathbf{k}} = (-1)^{k_i} \ket{H_\mathbf{k}}$. Hence, $\ket{H_\mathbf{k}}$ is an eigenstate of $S_i$ with eigenvalue $(-1)^{k_i}$.
 It is convenient to write arbitrary states in the hypergraph basis:
\begin{align}
    \rho = \sum_{\mathbf{k},\mathbf{k'}} c_{\mathbf{k},\mathbf{k'}} \dyad{H_\mathbf{k}}{H_\mathbf{k'}}. 
    \label{eq:gen_state}
\end{align}
Later we will purify states in this form to the state  $\ket{H_{\mathbf{0}}}$.

\subsection{Operations on Hypergraph States}
Many operations on hypergraph states can be represented in 
a graphical manner. In the following we explain the effect 
of applying Pauli gates $X$ and $Z$, measuring in the 
corresponding basis $\sigma_x$ and $\sigma_z$,  discuss 
how to represent the $\cnot$ gate graphically \cite{Mari_2017}, 
and introduce the \textit{reduction operator} $P_{v_1,v_2}$ 
which we will need later. Note that in the following for Pauli
matrices we use $X$ and $Z$ to denote the corresponding unitary
transformation and $\sigma_x$ and $\sigma_z$ to denote the measurements.
We only discuss transformations 
that are needed in the current paper, an overview on other
transformations can be found in 
Ref.~\cite{Mari_2019_Dissertation}.

\begin{figure}[t]
    \centering
    \includegraphics{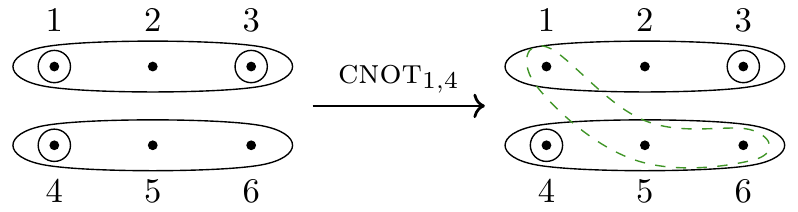}
    \caption{Example of a $\cnot_{1,4}$ gate (with control qubit 1 
    and target qubit 4) performed on a hypergraph state. 
    Left: Hypergraph with vertex set $V = \lbrace 1, \dots , 6 \rbrace$ 
    and edge set $E = \lbrace \lbrace 1 \rbrace, \lbrace 1,2,3 \rbrace, 
    \lbrace 3 \rbrace, \lbrace 4 \rbrace, \lbrace 4, 5, 6 \rbrace \rbrace$.  
    Right: Hypergraph after applying $\cnot_{1,4}$. A new edge $\lbrace 1,5,6 \rbrace $ 
    appeared while the edge $\lbrace 1 \rbrace$ vanished. The effect of 
    applying the $\cnot_{1,4}$ gate is to introduce or delete edges from 
    the set  $E_4 = \lbrace \lbrace 1 \rbrace, \lbrace 1,5,6 \rbrace \rbrace$. 
    The underlying rule is the following \cite{Mari_2017}: One takes
    the so-called adjacency $\AAA(4)$ of the target qubit $t=4$, where 
    one first considers all edges that contain $t$, but then removes $t$ 
    from it. Here, we have $\AAA(4)=\{\{\},\{5,6\}\}.$ Then, $E_4$ contains 
    all edges which are unions of edges from $\AAA(4)$ and the edge 
    $\lbrace 1 \rbrace$ of the control qubit $c=1$. 
     }
    \label{fig:hypergraph_cnot}
\end{figure}

We have already mentioned the action of the unitary transformation 
$Z_v$ on some qubit $v$. It adds the edge $e = \lbrace v \rbrace$ 
to the set of edges $E$, if it was not contained before, or removes 
it otherwise. For example applying $Z_2$ and $Z_3$ to the left 
hypergraph state in \cref{fig:hypergraph_cnot} would add a circle 
at vertex $2$ and remove the one at vertex $3$.

\begin{figure}
    \centering
    \includegraphics{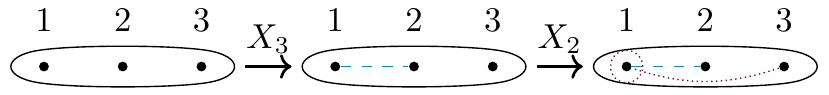}
    \caption{Application of $X$ operators on qubits 3 and 2. We first apply $X_3$ on the 
    left graph. The adjacency of qubit 3 is given by $\AAA(3) = \lbrace \lbrace 1,2 \rbrace \rbrace$. 
    This new edge is shown by the blue dashed line in the middle graph. We then 
    apply $X_2$ to the middle graph. The adjacency of qubit 2 is given by 
    $\AAA(2) = \lbrace \lbrace 1 \rbrace, \lbrace 1,3 \rbrace \rbrace$. These 
    new edges are shown by the dotted purple lines in the right graph.
    }
    \label{fig:apply_X}
\end{figure}

The unitary transformation $X_v$ on a vertex $v$ of a hypergraph state 
$\ket{H}$ corresponding to the hypergraph $H = (V,E)$ is given by
\begin{align}
    X_v \ket{H} 
    = \prod_{e \in E} C_e  \prod_{e' \in \AAA(v)} C_{e'}  \ket{+}^{\otimes \lvert V \rvert },
\end{align}
where $\AAA(v)$ is the \textit{adjacency} of vertex $v$. This is a set of edges defined as 
\begin{align}
\AAA(v) = \lbrace e \setminus \lbrace v \rbrace \mid e \in E \text{ with } v \in e \rbrace. 
\end{align}
In words, to build the adjacency $\AAA(v)$ one first takes the set of edges that contain $v$
and then removes $v$ from them. Examples of local transformations $X$ are given in 
\cref{fig:apply_X}.

Let us discuss now the graphical description of some local measurements 
on hypergraph states. In order to derive the post-measurement state after 
measuring vertex $v$, we can expand the state $\ket{H}$ at this vertex as
\begin{align}
\ket{H} = \frac{1}{\sqrt{2}} \left( \ket{0}_v \ket{H_0} \pm \frac{1}{\sqrt{2}} \ket{1}_v \ket{H_1} \right),
\end{align}
where $\ket{H_0}$ and $\ket{H_1}$ are new hypergraph states with vertex set $V_0 = V_1 = V \setminus v$ 
and edge sets $E_0 = \lbrace e \in E \mid v \notin e \rbrace$ and $E_1 = E_0 \cup \AAA(v)$
\cite{Mari_2019_Dissertation}. After measuring $\sigma_z$, we therefore either get the state 
$\ket{H_0}$ or $\ket{H_1}$. Measuring $\sigma_x$ leads to a superposition of these two states
and often the post-measurement state is then not a hypergraph sate anymore. In our case, 
we only measure $\sigma_x$ on qubits which are separated from other parts of the system. 
that is where $\ket{H_0} = \ket{H_1}$.

Applying a $\cnot_{ct}$ gate on a hypergraph state $H$, where $c$ is the control 
and $t$ the target, introduces or deletes hyperedges of the set 
$E_t = \lbrace e_t \cup {c} \mid e_t \in \AAA(t)  \rbrace $. The new edge set after 
applying  $\cnot_{ct}$ is given by 
\begin{align}
E' = E \triangle E_t,
\end{align}
where $A \triangle B = A \cup B \setminus A \cap B$ is the symmetric difference 
of two sets. Since $C_e^2 = \1$, double edges cancel out. Therefore, the operation 
$\cnot_{ct}$ deletes edges which are in $E$ and $E_t$ and introduces edges which 
are only in $E_t$. For example in the left part of \cref{fig:hypergraph_cnot}, 
the neighbourhood of vertex $4$ is given by $N(4) = \lbrace \{\} , \lbrace 5, 6 \rbrace \rbrace$ 
and therefore $E_4 = \lbrace \lbrace 1 \rbrace, \lbrace 1,5,6 \rbrace \rbrace$.

\begin{figure}
    \centering
    \includegraphics{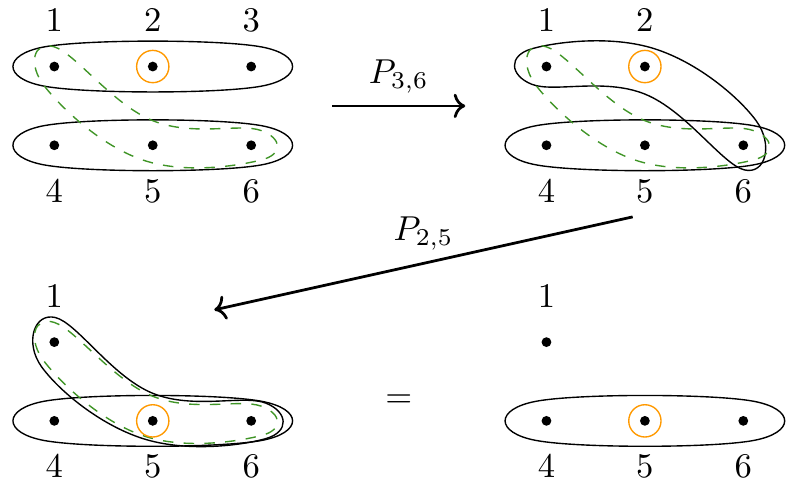}
    \caption{Application of the reduction projector $P_{3,6}$ and $P_{2,5}$. The projector 
    merges two vertices and its corresponding edges to one. In the first step, we merge 
    vertices 3 and 6. In the second step we merge vertices 2 and 5. This results in two times the same edge, the green dashed edge $\lbrace1,5,6 \rbrace$ and the edge which was 
    initially $\lbrace1,2,3 \rbrace$ and such double edges cancel out.
    }
    \label{fig:reduction_projector}
\end{figure}

Finally, another operator which will be important later is the \textit{reduction operator} 
$P_{v_1,v_2}$, which maps two qubits to a single qubit. In the computational basis, the 
reduction operator is written as 
\begin{align}
  P_{v_1,v_2} = \dyad{0}{00} + \dyad{1}{11}.  
\end{align}
It merges 
two vertices $v_1,v_2$ to one which we call $v_2$. This action changes edges which contain 
$v_1$ into edges which contain $v_2$ and deletes edges $e, e'$, with $e \neq e'$ 
but $(e \setminus \lbrace v_1 \rbrace) = (e' \setminus \lbrace v_2 \rbrace)$.
 The new edge set will therefore be 
 \begin{align}
     E' = (\{e \in E \vert v_1 \notin e \rbrace  
     \triangle \lbrace f \cup  \lbrace v_2 \rbrace \vert f \in \AAA(v_1)\rbrace ). \notag 
 \end{align}
An example is shown in \cref{fig:reduction_projector}.

\section{The CKDdV Purification Protocol} 
\label{sec:protocol}

In this section we discuss the only known protocol which works for 
hypergraph states \cite{Carle_2013}, we will refer to it as the CKDdV 
protocol. Originally, it was formulated for more general LME states.  
We first  reformulate the purification protocol in a graphical manner, 
which makes it intuitively understandable. Based on this reformulation,
we can then propose improvements.

In the simplest case, the aim is to purify a three-qubit state $\rho$ to 
a pure hypergraph state, chosen to be the state $\ket{H_{\mathbf{0}}} = C_{\lbrace 123 \rbrace } \ket{+}^{\otimes 3}$. The state is distributed between three parties, Alice, 
Bob, and Charlie. In the following, we explicitly describe the sub-protocol 
which reduces noise on Alice's qubit. There are equivalent sub-protocols 
on Bob's and Charlie's qubits.
The protocol is performed on two copies of a state $\rho$.  Alice holds qubit  $a_1$ 
of the first state and qubit $a_2$ of the second state, equivalently for Bob and Charlie.

The key idea of the protocol is to induce a transformation on the basis elements 
of the form
\begin{align}
\ket{H_{i, j, k}} \ket{H_{i', j', k'}} \rightarrow  \delta_{i, i'} \ket{H_{i, j+j', k+k'}}, \label{eq:transition1}
\end{align}
where $\delta_{i, i'}$ denotes the Kronecker delta. This means that the 
sub-protocol compares the indices $i,i'$ on Alice's qubits, and the state 
is discarded when $i \neq i'$. This map drives a general state as in 
Eq.~(\ref{eq:gen_state}) closer to the desired hypergraph state. In detail, 
the sub-protocol which implements this transition  is given by:


\begin{protocol}[CKDdV protocol] \label{prot:main} \text{ }

\noindent 
(0) Alice, Bob, and Charlie share two copies of a state.

\noindent
(i) Alice applies a local $\cnot_{a_1, a_2}$ gate on her qubits. 

\noindent
(ii) Bob and Charlie apply local reduction operators $P_{v_1,v_2}$ on their qubits. 

\noindent
(iii) Alice measures qubits $a_1$ in the $\sigma_x$ basis. She keeps the state, if the outcome is ``$+1$'', and discards it otherwise.
\end{protocol}

\begin{figure}
    \centering
    \includegraphics{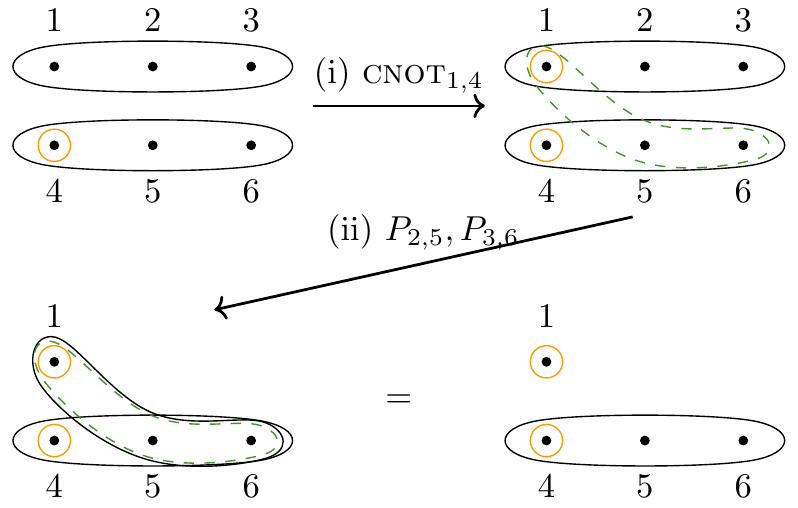}
    \caption{The CKDdV protocol, as described in in \cref{prot:main}. In the
    figure, the transformation of the two basis elements $\ket{H_{000}} \ket{H_{1 0 0}}$
    is shown.
    In step (i), Alice performs a local $\cnot_{1,4}$ gate. Then, Bob and Charlie 
    apply local reduction operators $P_{2,5}$ and $P_{3,6}$, respectively. Double 
    edges cancel out, so that the green dashed line and the former edge 
    $\lbrace1,2,3\rbrace$ vanish. In step (iii), Alice measures qubit 1 
    in the $\sigma_x$ basis. If there is a single-qubit edge on vertex 1, 
    as the orange one in this figure, her measurement outcome will be 
    ``$-1$'' and therefore the state gets discarded. If one ignores all 
    orange single-qubit edges in the figure, this corresponds to the 
    transformation of the basis elements $\ket{H_{000}} \ket{H_{0 0 0}}$. 
    In this case, Alice's measurement outcome will be ``$+1$'' and 
    the remaining state $\ket{H_{000}}$ is kept.
    }
    \label{fig:protocol}
\end{figure}

In \cref{fig:protocol} it is shown how the basis elements 
$\ket{H_{000}} \ket{H_{i 0 0}}$ transform.

In order to purify the full state, one needs to choose a 
sequence of sub-protocols in which these sub-protocols  are applied
on different parties. In Ref.~\cite{Carle_2013}, the sequence 
ABC-CAB-BCA was favoured, as it seems to perform better than just 
repeating the sequence ABC. The reason is  that the qubit of Charlie 
becomes more noisy due to the back action from the sub-protocols 
purifying Alice's and Bob's qubits.

\section{Improving the Protocol Performance} \label{sec:improvedprot}

In order to purify towards one state of a certain fidelity, one needs a number 
of input states, which depends exponentially on the number of iterations, as in 
each run of the protocol a certain fraction of states is discarded. Therefore 
it is of high interest to apply the subprotocols in a sequence which works 
as efficient as possible. As already pointed out by Carle \textit{et al.} 
\cite{Carle_2013}, it depends on the input state which sequence is the most 
advantageous and it is not trivial to see which sequence is optimal. 
Carle \textit{et al.}\ decided to use the sequence $S={\rm ABC-CAB-BCA}$ 
in all their applications, since it performs well in many cases. In the following
we will ask whether the proposed sequence really is the best and how we can  
potentially find better sequences.

One should notice that in step (ii) of the protocol a large fraction of states
is discarded. The operator $P_{v_1,v_2}$ corresponds to a positive map, which maps 
two qubits, which are in the same state, to one qubit and both qubits are discarded, 
if they are in different states. This can be seen as one outcome of a measurement. 
In the second part of this section we will ask whether one can reduce the amount 
of discarded states.

\subsection{Improved and Adaptive Sequences} \label{sec:sequenc}
Consider a noisy three-qubit state $\rho(p)$, where $p$ is a noise parameter 
for some noise model,  which should be purified to the pure hypergraph state 
$\dyad{H_{000}}$. Clearly, for a fixed sequence $S$ there is a maximal amount 
of noise until which the state can still be purified and there is a regime, where 
one cannot purify it any more. 

Interestingly, for some parameter regimes where the state cannot be purified, 
the purification protocol does not converge towards a state with random noise, 
but towards a specific state which is a mixture of two states: either 
$\frac{1}{2} (\dyad{H_{000}} + \dyad{H_{001}})$, 
$\frac{1}{2} (\dyad{H_{000}} + \dyad{H_{010}})$, or 
$\frac{1}{2} (\dyad{H_{000}} + \dyad{H_{100}})$. 
This observation gives insights about how good the purification works 
on different parties. The protocol eliminates noise on two parties but 
fails on the third party. For example if we apply sequence $S={\rm ABC}$, 
in the cases we tested, there is a regime, where the state does not get 
purified but converges to $\frac{1}{2} (\dyad{H_{000}} + \dyad{H_{001}})$. 

This is consistent with the explanation given in Ref.\ \cite{Carle_2013} 
that the purification has an disadvantage on Charlie's site. It may be
explained as follows: By performing the protocol at one party, one aims 
to reduce noise on this party. As an unwanted side effect, one increases 
noise on the other parties. This happens because if there is noise on 
the first input state, the local reduction operator will ``copy'' it to 
the second state (see \cref{eq:transition1}). 
So,  when choosing sequence  $S= {\rm ABC}$, one increases the noise on 
Charlie's qubit two times before purifying it the first time.

How well the protocol performs on each party can be analysed using 
the measurement statistics obtained in step (iii) of the protocol. 
The probability to measure outcome ``$+1$'' in step (iii) on a qubit 
belonging to a certain party gives insights, how much noise the state 
on this party has. On the perfect target state, one does not detect any 
noise and therefore measures outcome ``$+1$'' with probability equal to 
one. If one applies the protocol to the state 
$\frac{1}{2} (\dyad{H_{000}} + \dyad{H_{001}})$, however, one obtains  
outcome ``$+1$'' with a probability equal to one or 0.5, depending on 
which subprotocol was applied. If it was the subprotocol where 
Alice's or Bob's qubits  are measured in step (iii), the probability is equal to one. If it was the subprotocol where  Charlie's qubit 
was measured the probability is 0.5. So, by evaluating the probabilities to measure outcome ``$+1$'' in step (iii) of the protocol, one can 
analyse the efficiency of the sequence.

All in all, we use two approaches to find better sequences. The first 
approach is to find an optimal sequence, which allows a high noise 
tolerance and will be applied later without further observation of 
the statistics. The second approach uses two sequences where we 
switch from one to the other depending on the measurement outcomes 
during the process. The first approach helps to find sequences which 
are more efficient also for purification of states with a low 
noise level. The second approach gives a method to purify 
states which would not be purifyable otherwise.

We assume that we know how often we want to apply the protocol and therefore how many copies of the initial state are needed. We perform the first subprotocol on all copies and get a certain fraction of output states. We then perform the second subprotocol on all output states and so on. With this procedure, we do multiple measurements on copies of states such that we can approximate a probability from the frequency of a certain outcome. We further assume that we repeat this procedure several times, that is, we produce a certain number of copies of input states, purify them and start again from the beginning. In each run, we can vary the sequence, using what we have learned in the run before.
We restricted ourselves to sequences of length nine. 
The best sequence we find in this way we call $S_1$.

It is not known how the efficiency of the sequence depends on the state. Therefore, even if the state is known, one needs to sample which sequence works best. However, there are some observations which can be used to find better sequences. We notice that it is reasonable to consider permutations of A, B, C together.  We further notice, that the first party of the triple experiences the largest impact. It is also a good strategy to address the same party in two consecutive rounds, that is on the last position of one triple and on the first position of the following triple. 
The sequence proposed by Carle \textit{et al.}\ fulfills all mentioned properties and is in principle a good starting sequence. After gaining experience how efficient the sequence works for the given state, one can exchange few positions and evaluate its impact. 
We will see later (i.e. in \cref{tab:adaptive_seq,tab:results_more_qubits}) that the optimal sequences $S_1$ we found for different states, support our observations. Every triple is a permutation of A,B,C. We see that except of one case, the first position of one triple is equal to  either the first or the last position of the previous triple.
However, it consumes many states to find good sequences. Another strategy could be to estimate the
corresponding density matrix of the given state by local tomography and find the optimal sequence by simulations on the computer.

With the second approach, we give a way to purify states which can 
not be purified by sequence $S_1$ because their initial fidelity is 
slightly beyond the threshold. We start using sequence $S_1$ and 
switch to sequence $S_2$ depending on the measurement outcomes of 
step (iii). Our switching condition is the following: After each 
measurement of step (iii), we evaluate the probability to measure 
``$-1$'' for the given party.  Based on the last three probabilities 
associated to the same party, we take a decision to switch or not. 
For  $\vec{x}$ being the vector of this three probabilities, where 
$x_3$ is the newest probability, we switch, if the product of 
the vectors $\vec{a} \vec{x}$ exceed a bound $b$ where  $\vec{a}$ 
is a weight vector. 
In real applications we can not evaluate the probability. We suggest to count appearance of certain outcomes and estimate the probability from the frequency. 

\begin{table}
    \centering
    \begin{tabular}{l|c | c | c }
        & $\EE_{\text{wn}} (\rho,p)$ & $\EE_{\text{deph}} (\rho,p)$ & $\EE_{\text{depo}} (\rho,p)$ \\ \hline
        $S_1$ &  ABC-CBA-ABC & ABC-CBA-CBA & ABC-CAB-BCA \\
        $S_2$  & BAB-CAB-ABA & CCC-ACB-CBC & BBB-BCB-BBB-BAB     \\
        $\vec{a}  $ & $(0.33,0.35,0.32)$ & $(0.35,0.43,0.21)$ & $(0.35, 0.34, 0.31)$  \\
        $b$ & 0.35 & 0.39 & 0.44 \\
    \end{tabular}
    \caption{Sequences $S_1$, $S_2$, approximate weight vectors $\vec{a}$, and bounds $b$ for states with three kinds of noise. Explanation see text.}
    \label{tab:adaptive_seq}
\end{table}

To see the efficiency of our methods, we consider different noise models. 
We analyze the influence of global white noise described by the channel
\begin{align}
    \EE_{\text{wn}} (\rho,p) = p \rho + \frac{1-p}{2^n} \1,
\end{align}
where $n$ is the number of qubits. In this section, the number 
of states is $n=3$. We further analyse local noise channels given 
by $\EE (\rho,p) = \bigotimes_{i=1}^n \EE^i  (\rho,p)$, where 
$\EE^i$  is either the dephasing channel
\begin{align}
    \EE_{\text{deph}}^i (\rho,p) = p \rho + \frac{1-p}{2} (\rho + Z_i \rho Z_i)
\end{align}
or the depolarizing channel
\begin{align}
    \EE_{\text{depo}}^i (\rho,p) = p \rho + \frac{1-p}{4} (\rho + X_i \rho X_i + Y_i \rho Y_i + Z_i \rho Z_i).
\end{align}

The sequences, weight vectors and bounds we found to be optimal are given 
in  \cref{tab:adaptive_seq}. To compare the approaches, we give the noise 
thresholds found in Ref.\ \cite{Carle_2013}, obtained by our sequence $S_1$, 
and by the adaptive approach in \cref{tab:results}. The sequences we found 
are also better in other perspectives. If we apply the new sequences $S_1$  
nine rounds on given input states, we see that the output states have a 
higher fidelity then after purifying the same state nine rounds using the 
sequence given in Ref.\ \cite{Carle_2013}.

\begin{table}
    \centering
    \begin{tabular}{l|c | c | c } 
        & $p_{\text{min}}$ from \cite{Carle_2013} & $p_{\text{min}}$ from $S_1$ & $p_{\text{min}}$ from \\
        & & & adaptive protocol \\  \hline
        $\EE_{\text{wn}} (\rho,p)$ & 0.6007 & 0.5878 & 0.5876 \\
        $\EE_{\text{deph}} (\rho,p)$ & 0.8013 & 0.7803 & 0.7747 \\ 
        $\EE_{\text{depo}} (\rho,p)$  & 0.8136 & 0.8136 & 0.8132 
    \end{tabular}
    \caption{Noise thresholds $p_{\text{min}}$ reproduced from Ref.\ \cite{Carle_2013}, gained from our sequences $S_1$ (see \cref{tab:adaptive_seq}), and for the adaptive approach.
    In the case of $\EE_{\text{depo}} (\rho,p)$ we found that the sequence from Ref.\ \cite{Carle_2013} was already the best sequence of length 9. Therefore there is no improvement of $p_{\text{min}}$ in this case.
    }
    \label{tab:results}
\end{table}

\subsection{Recycling of Discarded States} \label{sec:Pprime}

If one wishes to purify a state using the CKDdV protocol one needs 
a high number of input states in order to obtain one state of a 
certain fidelity. Let us count how many states we need to have 
one state after applying the protocol once. In  step (0) of the 
protocol, one takes two input states. One does not loose states 
by applying $\cnot$ in step (i). By applying the reduction operator $P_{v_1,v_2}$, approximately $\frac{1}{2}$ of the pairs are lost. 
Since this operator is applied on two parties in step (ii), one 
needs approximately four pairs. In step (iii), one measures
outcome ``$+1$'' with a probability $\leqslant 1$. This probability 
depends on the fidelity of the states and increases with increasing 
fidelity. So, in total, approximately $8$ input states are 
required to obtain one output state. To prepare a state for which 
we need to apply the protocol $m$ times, we need more than $8^m$ 
input stats. To purify, for example, a state of initial fidelity 
0.93 to a state of fidelity of 0.994, we need three steps.
The required number of input states to obtain one output state is 
roughly $8.7^3 \approx 660$. If we want to purify the same state 
to a fidelity of  $0.999$, which we reach after six steps, we 
need about $8.38^6 \approx 346\ 000$ input states to get one 
new state.

It is natural to try to use the available quantum states more 
efficiently. In step (ii) of the CKDdV protocol, one performs 
a projective measurement and considers only one outcome, namely  $P_{v_1,v_2}$, which we get with probability approximately 
$\frac{1}{2}$. We suggest to use the states which were discarded 
because we measured something different than $P_{v_1,v_2} $.
The second reduction operator $P_{v_1,v_2}^\perp $ is 
perpendicular to $P_{v_1,v_2}$ and defined as
\begin{align}
    P_{v_1,v_2}^\perp  = \dyad{0}{10} + \dyad{1}{01} = P_{v_1,v_2} (X_{v_1} \otimes \1_{v_2}).
\end{align}
As $P_{v_1,v_2} $, the operator $P_{v_1,v_2}^\perp $ is a positive 
map. It maps two qubits, which are in different states, to one qubit. 
This can be seen as a different measurement outcome than $P_{v_1,v_2}$,
or one may interpret the set $\{P_{v_1,v_2}, P_{v_1,v_2}^\perp\}$ as a 
quantum instrument.

In the original CKDdV protocol one keeps the state only after 
measuring $P_{b_1,b_2} P_{c_1,c_2}$. There are three more possible 
measurement outcomes: $P_{b_1,b_2} P_{c_1,c_2}^\perp$, 
$P_{b_1,b_2}^\perp P_{c_1,c_2}$, and 
$P_{b_1,b_2}^\perp P_{c_1,c_2}^\perp$. In the cases of 
measuring $P_{v_1,v_2}^\perp $ on at least one party, one obtains 
a post measurement state on which one can apply some corrections to  
get a state, which is similar to the input state. One can collect 
these states and further purify them.

So, one can write down a modified protocol of the CKDdV protocol. 
Here, we give the sub-protocol which reduces noise on Alice's qubits. 
The sub-protocols for Bob and Charlie work equivalently.

\begin{protocol}[Improved CKDdV protocol] \text{ } \label{prot:new}

\noindent
(0) Alice, Bob, and Charlie share two copies of a state.

\noindent
(i) Alice applies a local $\cnot_{a_1, a_2}$ gate on her qubits. 

\noindent
(ii) Bob and Charlie perform a measurement on their qubits and  measure the local reduction operators $P_{v_1,v_2}$ and $P_{v_1,v_2}^\perp $.  If the measurement outcome for Bob and Charlie was $P_{v_1,v_2}$, continue with step (iiia). Else, continue with (iiib)

\noindent
(iiia) After Bob and Charlie both measured $P_{v_1,v_2}$, Alice measures qubits $a_1$ in the $\sigma_x$ basis. She keeps the state, if the outcome is ``$+1$'', and discards it otherwise.

\noindent
(iiib) After measuring $P_{v_1,v_2}^\perp$ on at least one pair of Bob and Charlie's qubits, Alice measures her qubit $a_1$ in the $\sigma_z$ basis. If she measure ``$+1$'', she keeps the state as it is. Otherwise, Bob and Charlie apply some local unitaries, which depend on the combinations of measurement outcomes in step (ii) and are given in \cref{tab:local_corrections}.
\end{protocol}

\begin{table}
    \centering
    \begin{tabular}{c|c | c}
       Measurement  & local correction & local correction \\ 
       outcomes & Bob & Charlie \\ \hline
       $P_{b_1,b_2} P_{c_1,c_2}^\perp$  & $Z$ & $\1$ \\
       $P_{b_1,b_2}^\perp P_{c_1,c_2}$  & $\1$ & $Z$ \\
       $P_{b_1,b_2}^\perp P_{c_1,c_2}^\perp$  & $Z$ & $Z$ \\
    \end{tabular}
    \caption{In \cref{prot:new} step (iiib), Alice measures her qubit $a_1$ in the $Z$ basis. If her outcome is ``$-1$'', Bob and Charlie have to apply local corrections to their qubits. The local corrections depend on their measurement outcomes from step (ii) and are given in this table. The first case is shown in \cref{fig:pprime}.
    }
    \label{tab:local_corrections}
\end{table}

The key idea is that output states from step (iiib) can be collected 
and further purified. In case of measuring $P_{v_1,v_2}^\perp $ on at least one party, the protocol gives us a transition 
\begin{align}
    \ket{H_{i, j, k}} \ket{H_{i', j', k'}} \rightarrow  \ket{H_{i', j+j', k+k'}}.  \label{eq:transition_new}
\end{align}
The resulting state has in general a lower fidelity than the input state. 
This is caused by the same reason of ``copying'' noise, as discussed before.
Since in the considered case the protocol does not reduce noise, the fidelity drops.  

\begin{figure}
    \centering
    \includegraphics{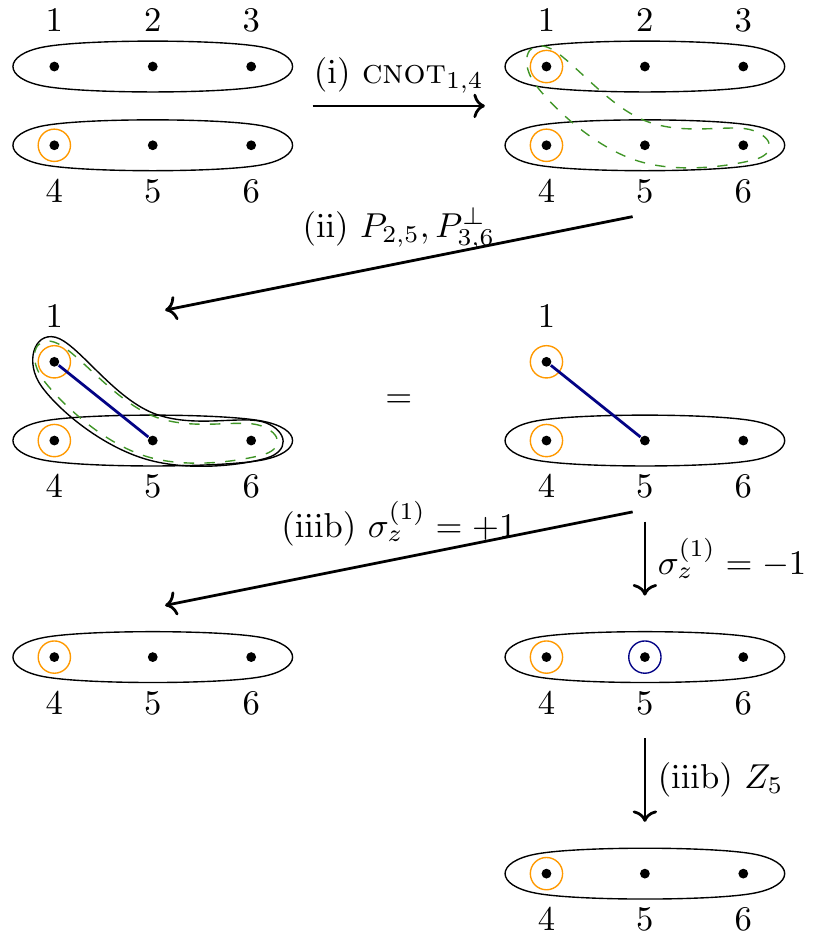}
    \caption{Modified \cref{prot:new} for the same initial states as shown in \cref{fig:protocol} for the case to measure $P_{b_1,b_2} P_{c_1,c_2}^\perp$ in step (ii). Alice performs a $\sigma_z^{(1)}$-measurement on her  qubit 1 of the state in the second raw. If she gets outcome ``$+1$'' in step (iiib), the resulting state is the same as the initial state (qubits 4, 5 and 6). If she  gets outcome ``$-1$'', Bob's qubit 5 has a decoration, which he needs to correct. After Bob applied a local $Z_5$ unitary on qubit 5, again the resulting state is the same as the initial state (qubits 4, 5 and 6). Note that this is only the case, if there is no noise on qubit 2 and 3, as shown in this figure. In general one obtains the state given in \cref{eq:transition_new}.
    }
    \label{fig:pprime}
\end{figure}

An example for \cref{prot:new} is shown in \cref{fig:pprime}, where we assume 
the case that Bob measures $P_{2,5}$  and Charlie measures $P_{3,6}^\perp$. In 
this case, the local correction after measuring outcome ``$-1$'' is applying a 
unitary $Z_5$ at qubit 5.

\begin{figure}
    \centering
    \includegraphics[width=1\linewidth]{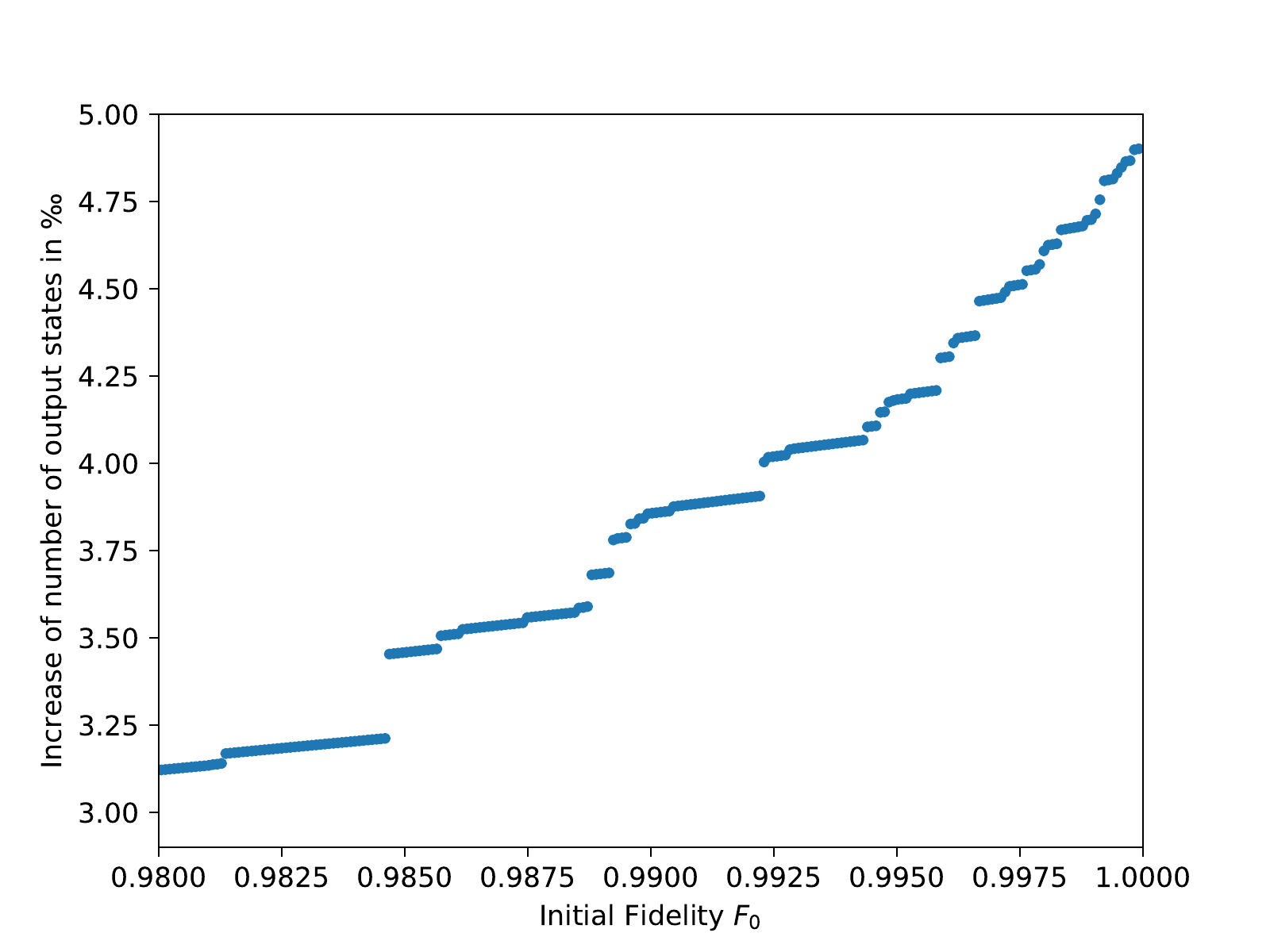}
    \caption{Effect of using \cref{prot:new} instead of the original CKDdV protocol.
    The input states are given by $\EE_{\text{wn}} (\dyad{H_{\mathbf{0}}},p)$. 
    We first apply \cref{prot:main} three times and computed the fidelity $F_3$ 
    of the output states. Then, we apply \cref{prot:new} on the same input 
    states and compare how many more output states of fidelity $\geqslant F_3$ 
    we get. The figure displays the increase of output states by using 
    \cref{prot:new}, depending on the fidelity $F_0$ of the input states.}
    \label{fig:reste}
\end{figure}

Given a certain number of input states which we want to purify to a target 
fidelity, we obtain more output states of the desired fidelity if we follow 
\cref{prot:new} instead of the original CKDdV protocol. The effect in the cases we tested turned out, however, to be small.  As input states, we chose the state $\dyad{H_{000}}$ mixed with white noise. We first applied \cref{prot:main} three 
times, that is, once on each party, and computed the fidelity $F_3$ of the output 
states. Then, we applied \cref{prot:new} on the same input states and compared how 
many more output states of fidelity $\geqslant F_3$ we get. In \cref{fig:reste} we show how much the number of output states increase by using \cref{prot:new}, depending on the fidelity $F_0$ of the input states. In the chosen cases, we get  approximately 
0.4 \%  more output states from using \cref{prot:new} instead of the CKDdV 
protocol. 

A similar idea of reusing states which get discarded in most protocols was proposed in Ref.\ \cite{Zhou_2020}. Zhou \textit{et al.}\ consider Bell states and reuse states from the last step of the protocol, which is equivalent to step (iii) in the CKDdV protocol.

\section{Generalisation to More Qubits}  \label{sec:morequbits}

\begin{figure}
    \centering
    \includegraphics{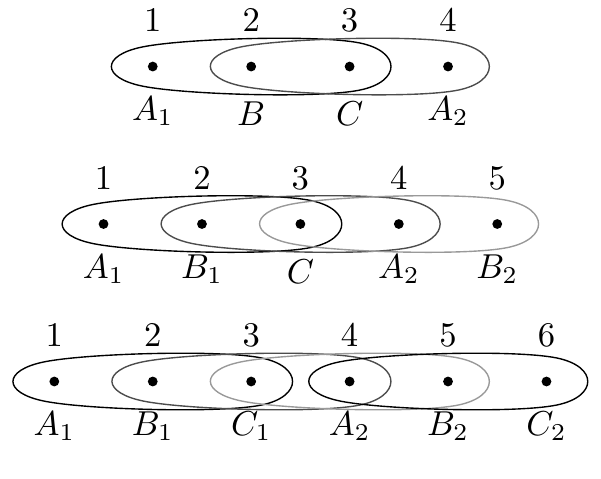}
    \caption{Linear 3-colorable and 3-regular hypergraph  states with 4, 5, and 6 qubits. The colors are denoted by $A$, $B$, and $C$. Note that two qubits which have the same color, for example qubits 1 and 4, still belong to different parties. Since we are restricted to local operations, we can only perform operations on qubits of the same party, that is in general not on qubits of the same color.}
    \label{fig:largerstates}
\end{figure}

The methods described here can also be applied to states with more 
qubits and different arrangement of edges. We restrict our attention 
to hypergraphs which are \textit{$k$-regular} and \textit{$k$-colorable}. 
A hypergraph is  $k$-regular, if  all edges $e\in E$ have order $k$ and
it is $k$-colorable, if it is possible to color vertices of a 
hypergraph using $k$ colors such that no two vertices of the same 
color share a common edge. 
For example, the hypergraph states shown in \cref{fig:hypergraph_cnot,fig:largerstates}  are 3-colorable and 3-regular. In this section we discuss purification protocols 
to hypergraph states of more than 3 qubits which are 3-colorable and 3-regular.
In the following, we will denote the colors by $A$, $B$, and $C$.

The protocols can be generalised by letting all parties holding qubits of color 
$A$ do what was described for Alice before. In the same way, parties holding 
a qubit of color $B$ or $C$ do what was described for Bob or Charlie, 
respectively. For a explicit formulation of the generalized protocol, see Ref.\ \cite{Carle_2013}.

\begin{table}
    \centering
    \begin{tabular}{l|c | c  | l} 
        & $p_{\text{min}}$ from & $p_{\text{min}}$  & sequence $S_1$\\
        & $S_{\text{CKDdV}}$  & from $S_1$ & \\ \hline
        $\EE_{\text{wn}} (\rho_3,p)$ & 0.6007 & 0.5878 & ABC-CBA-ABC \\
        $\EE_{\text{wn}} (\rho_4,p)$ &  0.4633  & 0.4396  & ABC-ACB-BCA \\
        $\EE_{\text{wn}} (\rho_5,p)$ &  0.3901  & 0.3486 & ABC-ABC-CBA \\
        $\EE_{\text{wn}} (\rho_6,p)$ &  0.3341  & 0.3017 & ABC-ACB-BAC* \\ \hline
        $\EE_{\text{deph}} (\rho_3,p)$ & 0.8013 & 0.7803  & ABC-CBA-CBA \\ 
        $\EE_{\text{deph}} (\rho_4,p)$ & 0.8014  & 0.7803 & ABC-CBA-CBA* \\
        $\EE_{\text{deph}} (\rho_5,p)$ & 0.8014  & 0.7803 & ABC-CBA-CBA* \\
        $\EE_{\text{deph}} (\rho_6,p)$ & 0.8014  & 0.7803 & ABC-CBA-CBA* \\ \hline
        $\EE_{\text{depo}} (\rho_3,p)$  & 0.8137 & 0.8136 & ABC-CAB-BCA  \\ 
        $\EE_{\text{depo}} (\rho_4,p)$ & 0.8306 & 0.8122 & BAC-CBA-CAB \\ 
        $\EE_{\text{depo}} (\rho_5,p)$ & 0.8358 & 0.8128 & ACB-BCA-CBA \\ 
        $\EE_{\text{depo}} (\rho_6,p)$ & 0.8144 & 0.8121 & ABC-CBA-CAB 
    \end{tabular}
    \caption{Noise thresholds $p_{\text{min}}$ for the sequence $S_{\text{CKDdV}}$ proposed in Ref.\ \cite{Carle_2013} and new sequences $S_1$. 
    The index of the state gives the number of qubits. In the case of $\EE_{\text{depo}} (\rho_3,p)$ we found that the sequence from Ref.\ \cite{Carle_2013} was already the best sequence of length 9. Therefore there is no improvement of $p_{\text{min}}$.
    When we found (non-trivially) different sequences of the same length, we marked them with an asterisk.  
    }
    \label{tab:results_more_qubits}
\end{table}

We analysed linear three-colorable states with up to six qubits under 
the influence of global white noise, dephasing and depolarisation. 
That is the states to which we want to purify are 
$U_{123}U_{234} \ket{+}^{\otimes 4}$, $U_{123}U_{234}U_{345} \ket{+}^{\otimes 5}$, 
and $U_{123}U_{234}U_{345}U_{456} \ket{+}^{\otimes 6}$, as shown in \cref{fig:largerstates}. We compare the noise threshold $p_{\text{min}}$ 
for the sequence proposed in  Ref.\ \cite{Carle_2013} with new sequences 
$S_1$, found using methods described in \cref{sec:sequenc}.

Our results are shown in \cref{tab:results_more_qubits}.
One sees that in the case of white noise for more qubits, the differences 
in the noise threshold  $p_{\text{min}}$ become more significant. Therefore, 
especially in these cases it is more relevant to find good sequences. 
For the tested states with dephasing and depolarisation noise, 
the noise threshold is constant or varies slightly, respectively.

\section{Conclusion and Outlook} \label{sec:conclusion}
In this paper we discussed protocols for entanglement purification of
hypergraph states. First, we reformulated the CKDdV protocol in a
graphical language. This offers a new way to understand the protocol,
furthermore, it allows to search for systematic extensions. 
Consequently, we introduced several improvements of the original 
protocol. These improvements are based on different sequences, 
adaptive schemes, as well as methods to recycle some of the unused 
states. While these modifications are conceptually interesting and 
can indeed improve the performance in various examples, the amount 
of the improvement in realistic examples seems rather modest.

The problem of finding efficient sequences is also relevant 
for purification protocols for other states and was raised 
for example in Ref.\ ~\cite{Aschauer_2005} in the context 
of two-colorable graph states. The methods developed here 
can be applied to this case, but also to all purification 
protocols which follow the concept  introduced by Bennett 
\textit{et al.}\  \cite{Bennett_1996}.

A further open question is how the effects of our methods scale 
with the number of qubits. Another open question is whether 
\cref{prot:new} can be further improved so that the effect 
gets more significant.

\section{Acknowledgments}
We thank Mariami Gachechiladze, Kiara Hansenne, Jan L.~B\"onsel, 
 Fabian Zickgraf, Yu-Bo Sheng, Lucas Tendick, and Owidiusz Makuta for discussions. This work was supported by 
the Deutsche Forschungsgemeinschaft  (DFG, German Research 
Foundation, project numbers 447948357 and 440958198), the 
Sino-German Center for Research Promotion (Project M-0294), 
the ERC (Consolidator Grant 683107/TempoQ), the German 
Ministry of Education and Research (Project QuKuK, BMBF Grant 
No. 16KIS1618K), and the Stiftung der Deutschen Wirtschaft.

\bibliographystyle{apsrev4-2}
\bibliography{references}

\end{document}